\documentclass[11pt,nofootinbib,floatfix,onecolumn,preprintnumbers]{revtex4}
\pdfoutput=1

\usepackage{amsmath,amssymb}
\usepackage{graphicx}
\usepackage{rotating}
\usepackage{color}
\usepackage{multirow} 
\usepackage{fontenc} 
\usepackage{slashed}
\usepackage{longtable}
\usepackage{hyperref}

\def\hc{\text{h.c.}}

\newcommand{\AddrBonn}{%
Bethe Center for Theoretical Physics and
Physikalisches Institut der Universit\"at Bonn,
Nussallee 12, D-53115 Bonn, Germany }

\newcommand{\AddrIFIC}{%
Instituto de F\'{\i}sica Corpuscular (CSIC-Universitat de Val\`{e}ncia),
Apdo. 22085, E-46071 Valencia, Spain
}

\preprint{IFIC/16-24}
\preprint{BONN-TH-2016-03}
\begin{document}

\title{Lepton Flavor Violation in the singlet-triplet scotogenic model}

\author{Paulina Rocha-Mor\'an}\email{procha@th.physik.uni-bonn.de}
\affiliation{\AddrIFIC}
\affiliation{\AddrBonn}

\author{Avelino Vicente}\email{avelino.vicente@ific.uv.es}
\affiliation{\AddrIFIC}

\begin{abstract}
We investigate lepton flavor violation (LFV) in the the
singlet-triplet scotogenic model in which neutrinos acquire non-zero
masses at the 1-loop level. In contrast to the most popular variant of
this setup, the singlet scotogenic model, this version includes a
triplet fermion as well as a triplet scalar, leading to a scenario
with a richer dark matter phenomenology. Taking into account results
from neutrino oscillation experiments, we explore some aspects of the
LFV phenomenology of the model. In particular, we study the relative
weight of the dipole operators with respect to other contributions to
the LFV amplitudes and determine the most constraining observables. We
show that in large portions of the parameter space, the most promising
experimental perspectives are found for LFV 3-body decays and for
coherent $\mu-e$ conversion in nuclei.
\end{abstract}

\maketitle

\section{Introduction}
\label{sec:intro}

Although the Standard Model (SM) of particle physics is supported by a
vast amount of experimental evidences, it is also known to be
incomplete due to its lack of solution for two central problems of
modern physics: neutrino masses and the dark matter (DM) of the
universe. Several SM extensions aiming at a common explanation for
these two issues have been put forward in recent years. The scotogenic
model, proposed by Ernest Ma in \cite{Ma:2006km}, constitutes one of
the most attractive proposals. In this model, the SM particle content
is enlarged with the introduction of a second scalar doublet and $N_N$
(with $N_N \geq 2$) singlet fermions, all charged under a
$\mathbb{Z}_2$ parity. This discrete symmetry forbids the usual
tree-level contribution to neutrino masses, which are induced at the
1-loop level, and gives rise to a stable state, a weakly-interacting
dark matter candidate. The phenomenology of this model has been
studied in great detail, see
\cite{Ma:2006fn,Kubo:2006yx,Hambye:2006zn,Sierra:2008wj,Suematsu:2009ww,Gelmini:2009xd,Adulpravitchai:2009gi,Aoki:2010tf,Ahn:2012cga,Schmidt:2012yg,Ma:2012if,Kashiwase:2012xd,Kashiwase:2013uy,Toma:2013zsa,Racker:2013lua,Klasen:2013jpa,Ho:2013hia,Ho:2013spa,Vicente:2014wga,Faisel:2014gda,Molinaro:2014lfa,Chowdhury:2015sla},
and several theoretical aspects have been discussed in the recent
literature, such as renormalization group running effects
\cite{Bouchand:2012dx,Merle:2015gea,Merle:2015ica} as well as new
model building directions
\cite{Ma:2008ym,Adulpravitchai:2009re,Ma:2013yga,Ma:2014eka,Yu:2016lof,Ahriche:2016cio}.

In this work we will concentrate on a simple extension~\footnote{See
  \cite{Restrepo:2013aga} for a general classification of scotogenic
  models leading to radiative neutrino masses and viable dark matter
  candidates.} of the minimal setup introduced in \cite{Ma:2006km}:
the singlet-triplet scotogenic model \cite{Hirsch:2013ola}. In this
variant of the scotogenic model, the fermion sector includes the
$SU(2)_L$ triplet $\Sigma$, which can mix with the singlet fermions
via the vacuum expectation value (VEV) of a real scalar, $\Omega$,
also triplet under $SU(2)_L$. The most relevant features of the
minimal model, radiative neutrino masses and a stable dark matter
candidate, are kept in this variant, while the singlet-triplet mixing
allows one to \emph{interpolate} between pure singlet DM
\cite{Ma:2006km} and pure triplet DM \cite{Ma:2008cu}, when the dark
matter candidate is fermionic. This leads to a richer phenomenology,
in particular to better prospects in direct DM detection experiments
\cite{Hirsch:2013ola}.

Lepton flavor violation (LFV) is one of the most important probes of
models with extended lepton sectors. In fact, precision high-intensity
experiments are sensitive to the existence of new physics at very high
energies, which makes flavor physics a powerful discovery tool, as
demonstrated by its central role in the making of the Standard
Model. Furthermore, very promising experimental projects in the search
for LFV will begin their operation in the near future. In addition to
the planned upgrade for the MEG experiment, which will improve its
sensitivity to $\mu \to e \gamma$ branching ratios as low as $6 \cdot
10^{-14}$ \cite{Baldini:2013ke}, other new experiments will also join
the effort. Among them, one can highlight the Mu3e
experiment~\cite{Blondel:2013ia}, which will look for the 3-body decay
$\mu \to 3 \, e$, as well as a plethora of experiments looking for
$\mu-e$ conversion in nuclei, like Mu2e
\cite{Carey:2008zz,Glenzinski:2010zz,Abrams:2012er}, DeeMe
\cite{Aoki:2010zz}, COMET \cite{Cui:2009zz,Kuno:2013mha} and
PRISM/PRIME \cite{Barlow:2011zza}, in all cases with spectacular
sensitivity improvements compared to previous experiments. This
remarkable multi-channel experimental effort in the search for LFV
encourages detailed LFV studies in specific neutrino mass models.

We study LFV in the singlet-triplet scotogenic model, in the spirit of
previous works for the singlet \cite{Toma:2013zsa} and triplet
\cite{Chao:2012sz} models~\footnote{See also \cite{Chowdhury:2015sla}
  for a general study of LFV in scotogenic models with higher
  $SU(2)_L$ representations.}. We will show that the model contains
large regions of parameter space with observable LFV rates and hence
will be probed in the near round of LFV experiments. Furthermore, we
will explore some aspects of the LFV phenomenology of the model, such
as the relative weight of the dipole operators with respect to other
contributions to the LFV amplitudes, and determine that the most
promising experimental perspectives are found for the LFV 3-body
decays $\mu \to 3 \, e$ and for coherent $\mu-e$ conversion in nuclei.

The rest of the paper is organized as follows: in Sec. \ref{sec:model}
we introduce the model whereas in Sec. \ref{sec:LFV} we review the
current experimental situation in the search for LFV and obtain
approximate expressions for the observables of
interest. Sec. \ref{sec:pheno} contains our phenomenological analysis
of the model. Finally, we summarize our results and draw our
conclusions in Sec. \ref{sec:conclusions} and present additional
analytical results in appendices \ref{app:LFVlag} and
\ref{app:LFVobs}.

\section{The model}
\label{sec:model}

We consider the singlet-triplet scotogenic model introduced in
\cite{Hirsch:2013ola}. The matter content of the model, as well as the
charge assignment under $SU(2)_L$, $U(1)_Y$ and $\mathbb{Z}_2$, is
shown in Table \ref{tab:MatterModel}. The quark sector, not included
in this table, is SM-like and has $\mathbb{Z}_2 = +1$. The new fields
beyond the SM particle content include two fermions: the singlet $N$
and the triplet $\Sigma$, both with vanishing hypercharge and odd
under the discrete $\mathbb{Z}_2$.  Regarding the new scalars, these
are the doublet $\eta$, also odd under $\mathbb{Z}_2$, and the real
triplet $\Omega$. The $SU(2)_L$ doublets $\phi$ and $\eta$ can be
decomposed as
\begin{equation}
\phi = \left( \begin{array}{c}
\phi^+ \\
\phi^0
\end{array} \right) \, , \quad \eta = \left( \begin{array}{c}
\eta^+ \\
\eta^0
\end{array} \right) \, ,
\end{equation}
and can be identified with the usual SM Higgs doublet and a new
\emph{inert} doublet. Regarding the $SU(2)_L$ triplets, $\Sigma$ and
$\Omega$, they are decomposed using the standard $2 \times 2$ matrix
notation as
\begin{equation}
\Sigma = \left( \begin{array}{cc}
\frac{\Sigma^0}{\sqrt{2}} & \Sigma^+ \\
\Sigma^- & -\frac{\Sigma^0}{\sqrt{2}}
\end{array} \right) \, , \quad \Omega = \left( \begin{array}{cc}
\frac{\Omega^0}{\sqrt{2}} & \Omega^+ \\
\Omega^- & -\frac{\Omega^0}{\sqrt{2}}
\end{array} \right) \, . \label{eq:triplets}
\end{equation}

\begin{table}[!t]
\centering
\begin{tabular}{|c||c|c|c||c|c||c|c|}
\hline
        & \multicolumn{3}{|c||}{Standard Model} &  \multicolumn{2}{|c||}{Fermions}  & \multicolumn{2}{|c|}{Scalars}  \\
        \cline{2-8}
        &  $L$  &  $e$  & $\phi$  & $\Sigma$ &  $N$   & $\eta$ & $\Omega$ \\
\hline     
generations & 3 & 3 & 1 & 1 & 1 & 1 & 1 \\                            
$SU(2)_L$ &  2    &  1    &    2    &     3    &  1   &    2   &    3     \\
$U(1)_Y$     & -1/2    &  -1    &    1/2    &     0    &  0   &    1/2   &    0  \\
$\mathbb{Z}_2$   &  $+$  &  $+$  &   $+$  &  $-$   & $-$  &  $-$  &  $+$    \\
\hline
\end{tabular}
\caption{Matter content and charge assignment of the singlet-triplet
  scotogenic model.}
\label{tab:MatterModel}
\end{table}

With the charge assignment in Table \ref{tab:MatterModel}, the most
general $\mathrm{SU(3)_c \otimes SU(2)_L \otimes U(1)_Y}$, Lorentz and
$\mathbb{Z}_2$ invariant Yukawa Lagrangian is given by
\begin{equation}
- \mathcal{L}_Y = Y_e^{\alpha \beta}\,\overline{L}_{\alpha} \, \phi \, e_{\beta} + Y_{N}^\alpha \, \overline{L}_{\alpha} \, \tilde{\eta} \, N + Y_{\Sigma}^\alpha \, \overline{L}_{\alpha} \, \tilde{\eta} \, \Sigma + Y_{\Omega} \, \overline{\Sigma} \, \Omega \, N + \hc \, . \label{eq:yukawa}
\end{equation}
Here we indicate the flavor indices $\alpha,\beta=1,2,3$ explicitly
and denote $\tilde{\eta} = i\sigma_2 \eta^{*}$, as usual. Gauge
contractions are omitted for the sake of clarity. The $\Sigma$ and $N$
fermions have Majorana mass terms,
\begin{equation}
- \mathcal{L}_M = 
\frac{1}{2} \, M_\Sigma \, \overline{\Sigma}^{c} \Sigma
+ \frac{1}{2} \, M_N \, \overline{N}^{c} N 
+ \hc \, . \label{eq:mass}
\end{equation}
Finally, the scalar potential can be written as~\footnote{The
  Lagrangian in Eqs. \eqref{eq:yukawa}, \eqref{eq:mass} and
  \eqref{eq:scpot} differs from the one in Ref.~\cite{Hirsch:2013ola}
  in two details: (i) some redundant terms in the scalar potential
  have been removed and the remaining ones have been renamed, and (ii)
  some couplings and mass terms have been normalized differently. The
  $SU(2)_L$ triplets $\Sigma$ and $\Omega$ also have a different
  normalization, see Eq.~\eqref{eq:triplets}.}
\begin{eqnarray}
\mathcal V &=& -m_{\phi}^2 \phi^\dagger \phi + m_{\eta}^2 \eta^\dagger \eta + \frac{\lambda_1}{2} \left( \phi^\dagger \phi \right)^2 + \frac{\lambda_2}{2} \left( \eta^\dagger \eta \right)^2 + \lambda_3 \left( \phi^\dagger \phi \right)\left( \eta^\dagger \eta \right) \nonumber \\ 
  &+& \lambda_4 \left( \phi^\dagger \eta \right)\left( \eta^\dagger \phi \right) + \frac{\lambda_5}{2} \left[ \left(\phi^\dagger \eta \right)^2 + \hc \right] - \frac{m_\Omega^2}{2} \, \Omega^\dagger \Omega  \nonumber \\
  &+& \frac{\lambda^{\Omega}_1}{2} \left( \phi^\dagger \phi \right) \left( \Omega^\dagger \Omega\right) + \frac{\lambda^{\Omega}_2}{4} \, (\Omega^\dagger \Omega )^2 + \frac{\lambda^{\eta}}{2} \left( \eta^\dagger \eta \right) \left( \Omega^\dagger \Omega\right) \nonumber \\
&+& \mu_1 \, \phi^\dagger \, \Omega \, \phi + \mu_2 \, \eta^\dagger \, \Omega \, \eta \, . \label{eq:scpot}
\end{eqnarray}

\subsection{Symmetry breaking and scalar sector}
\label{subsec:scalar}

We will assume that the scalar potential in Eq. \eqref{eq:scpot} is
such that
\begin{equation}
\langle \phi^0 \rangle = \frac{v_\phi}{\sqrt{2}} \, , \quad \langle \Omega^0 \rangle = v_\Omega \, , \quad \langle \eta^0 \rangle = 0 \, , \label{eq:vevs}
\end{equation}
with $v_\phi, v_\Omega \ne 0$. These vacuum expectation values (VEVs)
are determined by means of the minimization conditions
\begin{eqnarray}
t_\phi &=& -m_\phi^2 \, v_\phi + \frac{1}{2} \lambda_1 v_\phi^3 + \frac{1}{2} \lambda_1^\Omega v_\phi v_\Omega^2 - \frac{1}{\sqrt{2}} \, v_\phi v_\Omega \, \mu_1 = 0 \, , \label{eq:tad1} \\
t_\Omega &=& -m_\Omega^2 \, v_\Omega + \lambda_2^\Omega v_\Omega^3 + \frac{1}{2} \lambda_1^\Omega v_\phi^2 v_\Omega - \frac{1}{2 \sqrt{2}} v_\phi^2 \, \mu_1 = 0 \, , \label{eq:tad2}
\end{eqnarray}
where $t_i \equiv \frac{\partial \mathcal V}{\partial v_i}$ is the
tadpole of $v_i$. The VEVs $v_\phi$ and $v_\Omega$ break the
electroweak symmetry and induce masses for the gauge bosons,
\begin{eqnarray}
m_W^2 &=& \frac{1}{4} \, g^2 \left( v_\phi^2 + 4 \, v_\Omega^2 \right) \, , \label{eq:mW} \\
m_Z^2 &=& \frac{1}{4} \left(g^2 + g'^2 \right) v_\phi^2 \, .
\end{eqnarray}
We note that the triplet VEV $v_\Omega$ contributes to the $W$ boson
mass, thus receiving constraints from electroweak precision tests. We
estimate that this VEV cannot be larger than about $4.5$ GeV (at $3
\sigma$).

The scalar spectrum of the model contains the $\mathbb{Z}_2$-even
scalars $\phi^0$, $\Omega^0$, $\phi^\pm$ and $\Omega^\pm$, and the
$\mathbb{Z}_2$-odd scalars $\eta^0$ and $\eta^\pm$. In the basis
$\text{Re} \left( \phi^0\, ,\, \Omega^0 \right)$, the mass matrix for
the $\mathbb{Z}_2$-even neutral scalars is given by
\begin{eqnarray}
\mathcal{M}_S^2 &=& \left(\begin{array}{cc}
-m_\phi^2 + \frac{3}{2} \lambda_1 v_\phi^2 + \frac{1}{2} \lambda_1^\Omega v_\Omega^2 - \frac{1}{\sqrt{2}} v_\Omega \, \mu_1
& \lambda^\Omega_1 v_\phi v_\Omega - \frac{1}{\sqrt{2}} v_\phi \, \mu_1 \\
\lambda^\Omega_1 v_\phi v_\Omega - \frac{1}{\sqrt{2}} v_\phi \, \mu_1
& -m_\Omega^2 + \frac{1}{2} \lambda^\Omega_1 v_\phi^2 + 3 \lambda^\Omega_2 v_\Omega^2
\end{array}\right) \, .
\end{eqnarray}
The lightest of the $S$ mass eigenstates, $S_1 \equiv h$, can be
identified with the SM Higgs boson with a mass $m_h \simeq 126$ GeV
recently discovered at the LHC, whereas the heaviest mass eigenstate,
$S_2$, is a new heavy Higgs boson not present in the SM. Regarding the
$\mathbb{Z}_2$-even charged scalars, their mass matrix in the basis
$\left( \phi^\pm\, ,\, \Omega^\pm \right)$ takes the form
\begin{eqnarray}
\mathcal{M}_{H^\pm}^2 &=& \left(\begin{array}{cc}
-m_\phi^2 + \frac{1}{2} \lambda_1 v_\phi^2 + \frac{1}{2} \lambda^\Omega_1 v_\Omega^2 + \frac{1}{\sqrt{2}} v_\Omega \, \mu_1 + \frac{1}{4} g^2 v_\phi^2 \xi_{W^\pm} 
& \frac{1}{\sqrt{2}} v_\phi \, \mu_1 - \frac{1}{2} g^2 v_\phi v_\Omega \xi_{W^\pm} \\
\frac{1}{\sqrt{2}} v_\phi \, \mu_1 - \frac{1}{2} g^2 v_\phi v_\Omega \xi_{W^\pm}
& -m_\Omega^2 + \frac{1}{2} \lambda^\Omega_1 v_\phi^2 + \lambda^\Omega_2 v_\Omega^2 + g^2 v_\Omega^2 \xi_{W^\pm}
\end{array}\right) \, . \nonumber \\
\end{eqnarray}
One of the $H^\pm$ mass eigenstates is the Goldstone boson that
becomes the longitudinal component of the $W$ boson, whereas the other
is a physical charged scalar. In what concerns the $\mathbb{Z}_2$-odd
scalars $\eta^{0,\pm}$, we first express the neutral $\eta^0$ field in
terms of its CP-even and CP-odd components as
\begin{equation}
\eta^0 = \frac{1}{\sqrt{2}} \left( \eta^R + i \, \eta^I\right) \, .
\label{eq:defetaRI}
\end{equation}
The conservation of the $\mathbb{Z}_2$ symmetry implies that the
$\eta^{R,I,\pm}$ fields do not mix with the rest of scalars. Their
masses are given by~\footnote{Although we provide analytical
  expressions for the masses in full generality, our analysis will
  assume CP conservation in the scalar sector, allowing us to consider
  that $\eta^R$ and $\eta^I$ do not mix.}
\begin{eqnarray}
m_{\eta^R}^2 &=& m_{\eta}^2 + \frac{1}{2}\left(\lambda_3 + \lambda_4 + \lambda_5 \right) v_\phi^2 + \frac{1}{2}\lambda^\eta v_\Omega^2 - \frac{1}{\sqrt{2}} \, v_\Omega \, \mu_2 \, \\
m_{\eta^I}^2 &=& m_{\eta}^2 + \frac{1}{2}\left(\lambda_3 + \lambda_4 - \lambda_5 \right) v_\phi^2 + \frac{1}{2}\lambda^\eta v_\Omega^2 - \frac{1}{\sqrt{2}} \, v_\Omega \, \mu_2 \, \\
m_{\eta^{\pm}}^2 &=& m_{\eta}^2 + \frac{1}{2}\lambda_3 v_\phi^2 + \frac{1}{2}\lambda^\eta v_\Omega^2 + \frac{1}{\sqrt{2}} \, v_\Omega \, \mu_2 \, .
\end{eqnarray}
We point out that the mass difference between the neutral $\eta$
scalars is controlled by the $\lambda_5$ coupling,
$m_{\eta^R}^2-m_{\eta^I}^2 = \lambda_5 \, v_\phi^2$, and thus vanishes
if $\lambda_5 = 0$. This will be relevant for the generation of
neutrino masses, as shown in Sec.~\ref{subsec:numass}.

Finally, we emphasize that the vacuum in Eq. \eqref{eq:vevs} breaks
$SU(2)_L \otimes U(1)_Y \to U(1)_Q$ but preserves the $\mathbb{Z}_2$
discrete symmetry. As we will discuss below, this gives rise to the
existence of a stable neutral particle which may play the role of the
dark matter of the universe.

\subsection{Neutrino masses}
\label{subsec:numass}

Before discussing neutrino masses we must comment on the
$\mathbb{Z}_2$-odd neutral fermions. The $\mathbb{Z}_2$-odd fields
$\Sigma^0$ and $N$ get mixed by the Yukawa coupling $Y_\Omega$ and the
non-zero VEV $v_\Omega$. In the basis $\left( \Sigma^0, N \right)$,
their Majorana mass matrix takes the form
\begin{equation}
\mathcal{M}_\chi = \left(\begin{array}{cc} M_\Sigma & Y_\Omega v_\Omega \\ 
Y_\Omega v_\Omega & M_N \end{array}\right) \, .
\end{equation}
The mass eigenstates $\chi_{1,2}$ are determined by the $2\times2$
orthogonal matrix $V(\alpha)$,
\begin{equation}
\left(\begin{array}{c}\chi_1\\ \chi_2\end{array}\right) = \left( \begin{array}{cc}
\cos \alpha & \sin \alpha \\
-\sin \alpha & \cos \alpha
\end{array} \right) \, \left(\begin{array}{c} \Sigma^0\\ N\end{array}\right) = V(\alpha)\left(\begin{array}{c} \Sigma^0\\ N\end{array}\right),
\end{equation}
such that
\begin{equation}
\tan(2\alpha) = \frac{2 \, Y_\Omega v_\Omega}{M_\Sigma - M_N} \, .
\end{equation}

\begin{figure}[t]
\centering
\includegraphics[scale=0.45]{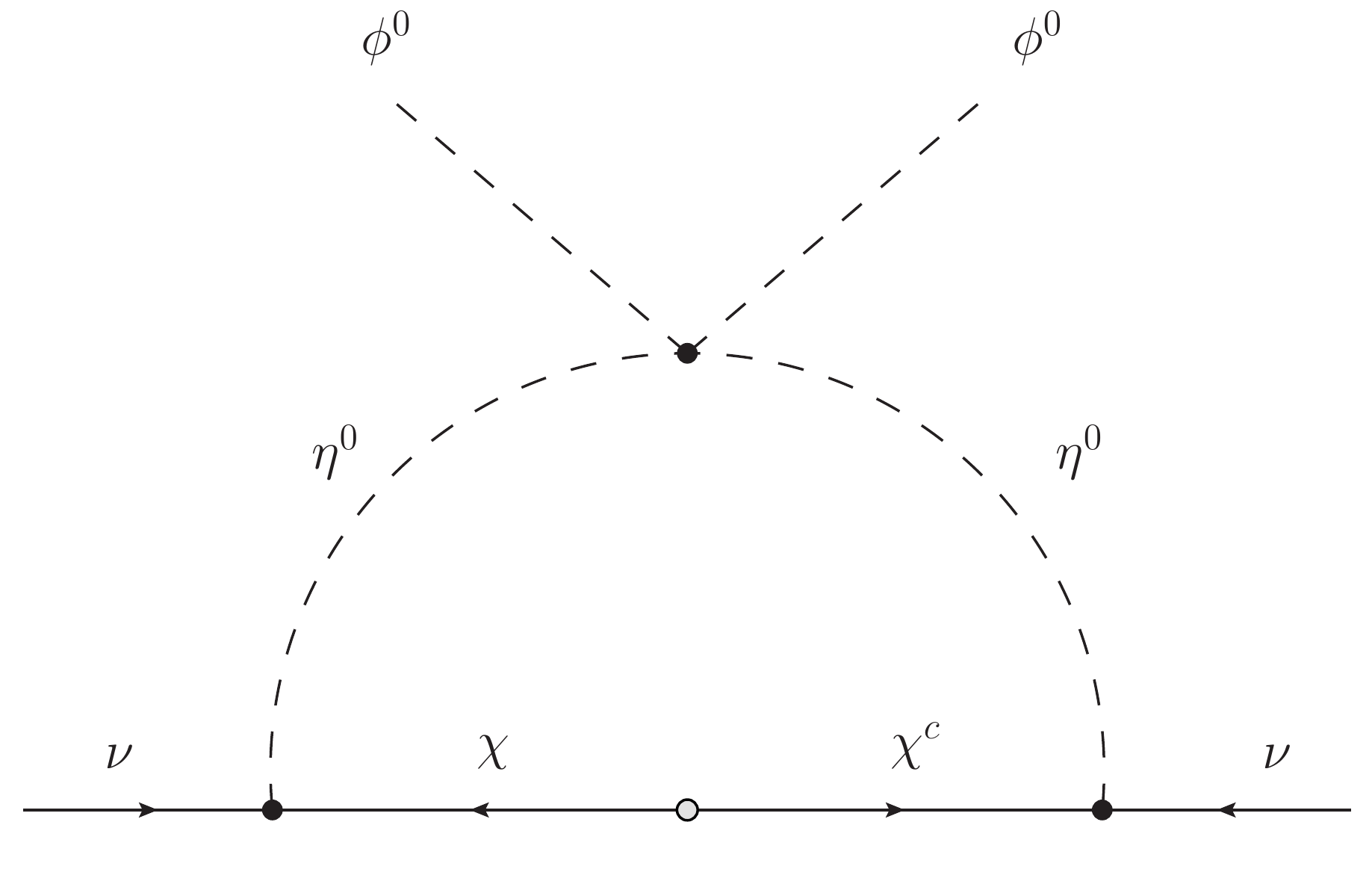}
\caption{1-loop neutrino masses in the singlet-triplet scotogenic
  model. Here $\eta^0 \equiv \left( \eta^R, \eta^I \right)$ and $\chi
  \equiv \left( \chi_1, \chi_2 \right)$.}
\label{fig:numass}
\end{figure}

The singlet-triplet scotogenic model generates Majorana neutrino
masses at the 1-loop level. This is shown in Fig.~\ref{fig:numass},
which actually includes four loop diagrams, since $\eta^0 \equiv
\left( \eta^R, \eta^I \right)$ and $\chi \equiv \left( \chi_1, \chi_2
\right)$. The resulting neutrino mass matrix can be written
as~\footnote{We correct this expression by including a factor of $1/2$
  missing in \cite{Hirsch:2013ola}.}
\begin{eqnarray}
(\mathcal{M}_\nu)_{\alpha\beta}&=&\sum_{\sigma=1}^2\left(\frac{ih_{\alpha \sigma}}{\sqrt{2}}\right)\left(\frac{-ih_{\beta \sigma}}{\sqrt{2}}\right)\left[I(M_{\chi_\sigma}^2,m_{\eta^R}^2)-I(M_{\chi_\sigma}^2,m_{\eta^I}^2)\right] \nonumber \\
&=&\sum_{\sigma=1}^2 \frac{h_{\alpha \sigma} \, h_{\beta \sigma} \, M_{\chi_\sigma}}{2 \, (4\pi)^2} \left[\frac{m_{\eta^R}^2\ln\left(\frac{M_{\chi_\sigma}^2}{m_{\eta^R}^2}\right)}{M_{\chi_\sigma}^2-m_{\eta^R}^2}
-\frac{m_{\eta^I}^2\ln\left(\frac{M_{\chi_\sigma}^2}{m_{\eta^I}^2}\right)}{M_{\chi_\sigma}^2-m_{\eta^I}^2}\right] \, , \label{eq:mnu}
\end{eqnarray}
where $h$ is a $3 \times 2$ matrix defined as
\begin{equation}
h=\left(\begin{array}{cc} 
\frac{Y_\Sigma^1}{\sqrt{2}} & Y_N^1 \\
\frac{Y_\Sigma^2}{\sqrt{2}} & Y_N^2 \\
\frac{Y_\Sigma^3}{\sqrt{2}} & Y_N^3
\end{array}\right) \cdot V^T(\alpha) \, ,
\end{equation}
and $I(m_1^2,m_2^2)$ is a Passarino-Veltman function evaluated in the
limit of zero external momentum. We note that $m_{\eta^R}^2 =
m_{\eta^I}^2$ leads to vanishing neutrino masses due to an exact
cancellation between the $\eta^R$ and $\eta^I$ loops. This was indeed
expected, since $m_{\eta^R}^2 = m_{\eta^I}^2$ implies $\lambda_5 = 0$
and a definition of a conserved lepton number would be possible in
this case. Furthermore, this justifies the choice $\lambda_5 \ll 1$,
which is natural in the sense of 't Hooft~\cite{'tHooft:1979bh}, given
that the limit $\lambda_5 \to 0$ increases the symmetry of the model.

It proves convenient to write the neutrino mass matrix in
Eq.~\eqref{eq:mnu} as
\begin{equation} \label{eq:mnumat}
\mathcal{M}_\nu = h \, \Lambda \, h^T \, ,
\end{equation}
where
\begin{equation}
\Lambda = \left( \begin{array}{cc}
\Lambda_1 & 0 \\
0 & \Lambda_2
\end{array} \right) \, , \quad \Lambda_\sigma = \frac{M_{\chi_\sigma}}{2 \, (4\pi)^2} \left[\frac{m_{\eta^R}^2\ln\left(\frac{M_{\chi_\sigma}^2}{m_{\eta^R}^2}\right)}{M_{\chi_\sigma}^2-m_{\eta^R}^2}
-\frac{m_{\eta^I}^2\ln\left(\frac{M_{\chi_\sigma}^2}{m_{\eta^I}^2}\right)}{M_{\chi_\sigma}^2-m_{\eta^I}^2}\right] \, .
\end{equation}
A neutrino mass matrix as the one in Eq.~\eqref{eq:mnumat} formally
resembles that obtained in the standard type-I seesaw with two
generations of right-handed neutrinos. In this case we can make use of
an adapted Casas-Ibarra parameterization
\cite{Casas:2001sr,Ibarra:2003up} to obtain an expression for the
Yukawa matrix $h$,
\begin{equation} \label{eq:CI}
h = U^\ast \, \sqrt{\widehat{\mathcal{M}}_\nu} \, R \, \sqrt{\Lambda}^{-1} \, .
\end{equation}
Here $R$ is a $3 \times 2$ complex matrix such that $R R^T =
\mathbb{I}_{3}$, where $\mathbb{I}_{3}$ is the $3 \times 3$ unit
matrix, and the neutrino mass matrix is diagonalized as
\begin{equation}
U^{T} \, \mathcal{M}_\nu \, U=\widehat{\mathcal{M}}_\nu\equiv
\left(
\begin{array}{ccc}
m_1 & 0 & 0\\
0 & m_2 & 0\\
0 & 0 & m_3
\end{array}
\right) \, ,
\label{eq:mnudiag}
\end{equation}
where 
\begin{equation}
\label{eq:PMNS}
U=
\left(
\begin{array}{ccc}
 c_{12}c_{13} & s_{12}c_{13}  & s_{13}e^{i\delta}  \\
-s_{12}c_{23}-c_{12}s_{23}s_{13}e^{-i\delta}  & 
c_{12}c_{23}-s_{12}s_{23}s_{13}e^{-i\delta}  & s_{23}c_{13}  \\
s_{12}s_{23}-c_{12}c_{23}s_{13}e^{-i\delta}  & 
-c_{12}s_{23}-s_{12}c_{23}s_{13}e^{-i\delta}  & c_{23}c_{13}  
\end{array}
\right)
\end{equation}
is the PMNS (Pontecorvo-Maki-Nakagawa-Sakata) matrix. Here $c_{ij} =
\cos \theta_{ij}$, $s_{ij} = \sin \theta_{ij}$ and $\delta$ is the
CP-violating Dirac phase~\footnote{In general, Eq. \eqref{eq:PMNS}
  could also include an additional Majorana phase. However, this will
  not be considered in this paper.}. Similarly to the type-I seesaw
with two right-handed neutrinos, the singlet-triplet scotogenic model
predicts a vanishing mass for the lightest neutrino. It has, however,
enough freedom to accommodate both neutrino spectra, Normal Hierarchy
(NH) and Inverted Hierarchy (IH), and the form of the complex $R$
matrix introduced in Eq.~\eqref{eq:CI} depends on this
choice~\cite{Ibarra:2003up},
\begin{eqnarray}
R &=& \left( \begin{array}{cc} 0 & 0 \\ \cos \gamma & \sin \gamma
  \\ -\sin \gamma & \cos \gamma
\end{array} \right) \quad \text{\bf for NH} \quad (m_1 = 0) \, , \\
R &=& \left( \begin{array}{cc}
\cos \gamma & \sin \gamma \\
-\sin \gamma & \cos \gamma \\
0 & 0
\end{array} \right) \quad \text{\bf for IH} \quad (m_3 = 0) \, .
\end{eqnarray}
We can finally make use of the previous expressions and write the
Yukawa couplings $h$ in terms of the PMNS matrix $U$, the eigenvalues
$m_i$ and the complex angle $\gamma$. In case of NH, one obtains
\begin{eqnarray}
h_{\alpha 1} &=& \frac{1}{\sqrt{\Lambda_1}} \left( \cos \gamma \, \sqrt{m_2} \, U_{\alpha 2}^\ast - \sin \gamma \, \sqrt{m_3} \, U_{\alpha 3}^\ast \right) \, , \label{eq:hfirst} \\
h_{\alpha 2} &=& \frac{1}{\sqrt{\Lambda_2}} \left( \sin \gamma \, \sqrt{m_2} \, U_{\alpha 2}^\ast + \cos \gamma \, \sqrt{m_3} \, U_{\alpha 3}^\ast \right) \, ,
\end{eqnarray}
whereas for IH one finds
\begin{eqnarray}
h_{\alpha 1} &=& \frac{1}{\sqrt{\Lambda_1}} \left( \cos \gamma \,
\sqrt{m_1} \, U_{\alpha 1}^\ast - \sin \gamma \, \sqrt{m_2} \,
U_{\alpha 2}^\ast \right) \, ,\\ h_{\alpha 2} &=&
\frac{1}{\sqrt{\Lambda_2}} \left( \sin \gamma \, \sqrt{m_1} \,
U_{\alpha 1}^\ast + \cos \gamma \, \sqrt{m_2} \, U_{\alpha 2}^\ast
\right) \, . \label{eq:hlast}
\end{eqnarray}

\subsection{Dark matter}
\label{subsec:DM}

The lightest state charged under the conserved $\mathbb{Z}_2$ parity
is stable and hence, if electrically neutral, it constitutes a
standard weakly-interacting dark matter candidate. Therefore, in what
concerns dark matter, the singlet-triplet scotogenic model contains
two distinct scenarios: (i) scalar dark matter, when the candidate is
the lightest neutral $\eta$ state, $\eta_R$ or $\eta_I$, and (ii)
fermion dark matter, when the candidate is $\chi_1$, the lightest
$\chi$ state. Even though we will not be concerned about dark matter
in this paper, we find it worth summarizing the main features of these
two scenarios:

\begin{itemize}

\item {\bf Scalar dark matter:} In this case the dark matter
  phenomenology resembles that of the inert doublet model
  \cite{Deshpande:1977rw} (see also
  \cite{Diaz:2015pyv,Queiroz:2015utg,Garcia-Cely:2015khw} for some
  recent works on dark matter in the inert doublet model).  Since in
  this scenario dark matter production in the early universe is driven
  by gauge interactions, there is no direct relation with LFV (driven
  by Yukawa interactions).

\item {\bf Fermion dark matter:} This scenario presents some of the
  most interesting features of the singlet-triplet scotogenic model
  \cite{Hirsch:2013ola}. The phenomenology dramatically depends on the
  nature of the dark matter candidate. In the two extreme cases this
  can be a pure $SU(2)_L$ singlet (when $\chi_1 \equiv N$) or a pure
  $SU(2)_L$ triplet (when $\chi_1 \equiv \Sigma$), while in general it
  will be an admixture of these two gauge eigenstates. When $\chi_1$
  is mostly singlet, the dark matter phenomenology is determined by
  Yukawa interactions and one expects a direct link between dark
  matter and LFV, as in the minimal scotogenic model
  \cite{Vicente:2014wga}. In contrast, the DM phenomenology of a
  mostly triplet dark matter candidate is driven by the known gauge
  interactions. This case has little impact on LFV and predicts a dark
  matter candidate with a mass of about $\sim 2.7$ TeV in order to
  reproduce the observed dark matter relic density. The parameter
  $Y_\Omega$, which determines the $N-\Sigma$ mixing, interpolates
  between these two cases, in a way completely analogous to DM in
  R-parity conserving supersymmetry.

\end{itemize}

\section{LFV observables}
\label{sec:LFV}

\subsection{Current experimental situation and future projects}
\label{subsec:expLFV}

No observation of a flavor violating process involving charged leptons
has ever been made. This has been used by many experiments to set
strong limits on the most relevant LFV observables, usually translated
into stringent bounds on the parameter space of many new physics
models. In what concerns the radiative decay $\ell_\alpha \to
\ell_\beta \gamma$, the experimental search is led by the MEG
collaboration. This experiment searches for the process $\mu \to e
\gamma$ and recently announced the limit $\text{BR}(\mu \to e \gamma)
< 5.7 \cdot 10^{-13}$ \cite{Adam:2013mnn}, about four times more
stringent than the previous bound obtained by the same
collaboration. The 3-body LFV decay $\mu \to 3 \, e$ was also searched
for long ago by the SINDRUM experiment \cite{Bellgardt:1987du}, which
obtained the limit $\text{BR}(\mu \to 3 \, e) < 1.0 \cdot 10^{-12}$,
still not improved by any experiment after almost 30 years. Another
$\mu-e$ LFV process of interest due to the existing bounds is $\mu-e$
conversion in nuclei. Among the experiments involved in this search we
may mention SINDRUM II, which searched for $\mu-e$ conversion in
muonic gold and obtained the impressive limit CR($\mu- e, {\rm Au}$)
$< 7\times 10^{-13}$~\cite{Bertl:2006up}. Finally, the current
experimental limits for $\tau$ lepton observables are less stringent,
with branching ratios bounded to be below $\sim 10^{-8}$.

In addition to the active LFV searches, some of them with planned
upgrades, several promising upcoming experiments will join the effort
in the next few years~\footnote{See
  \cite{Bernstein:2013hba,Mihara:2013zna,Signorelli:2013kla} for
  recent reviews.}. The MEG collaboration has announced plans for
upgrades which will allow this experiment to reach a sensitivity to
branching ratios as low as $6 \cdot 10^{-14}$
\cite{Baldini:2013ke}. Significant improvements are also expected for
$\tau$ observables from searches in B factories
\cite{Aushev:2010bq,Bevan:2014iga}, although the expected
sensitivities are still less spectacular than those for $\mu$
observables. Regarding the new projects, the most promising ones are
expected in searches for $\mu \to 3 \, e$ and $\mu-e$ conversion in
nuclei.  The Mu3e experiment, which plans to start data taking soon,
announces a sensitivity for $\mu \to 3 \, e$ branching ratios of the
order of $\sim 10^{-16}$ \cite{Blondel:2013ia}. In case no discovery
is made, this would imply an impressive improvement of the current
bound by $4$ orders of magnitude. Regarding $\mu-e$ conversion in
nuclei, the competition will be shared by several experiments, with
expected sensitivities for the conversion rate ranging from $10^{-14}$
to an impressive $10^{-18}$. These include Mu2e
\cite{Carey:2008zz,Glenzinski:2010zz,Abrams:2012er}, DeeMe
\cite{Aoki:2010zz}, COMET \cite{Cui:2009zz,Kuno:2013mha} and, in the
long term, the future PRISM/PRIME \cite{Barlow:2011zza}.

Finally, even though in this paper we concentrate on low-energy
processes, we emphasize that colliders can also play a relevant role
in the search for LFV. For instance, there is currently an intriguing
hint at CMS for Higgs boson LFV decays into $\tau \mu$
\cite{Khachatryan:2015kon}. This anomaly seems to require an
explanation based on an extended scalar sector (see
e.g. \cite{Sierra:2014nqa,Dorsner:2015mja}), in principle not related
to the problem of neutrino masses, and cannot be accommodated in the
model under investigation. For reference, in Tab.~\ref{tab:sensi} we
collect present bounds and expected sensitivities for the most popular
low-energy LFV observables.

\begin{table}[tb!]
\centering
\begin{tabular}{|c|c|c|}
\hline
LFV Process & Present Bound & Future Sensitivity  \\
\hline
    $\mu \rightarrow  e \gamma$ & $5.7\times 10^{-13}$~\cite{Adam:2013mnn}  & $6\times 10^{-14}$~\cite{Baldini:2013ke} \\
    $\tau \to e \gamma$ & $3.3 \times 10^{-8}$~\cite{Aubert:2009ag}& $ \sim3\times10^{-9}$~\cite{Aushev:2010bq}\\
    $\tau \to \mu \gamma$ & $4.4 \times 10^{-8}$~\cite{Aubert:2009ag}& $ \sim3\times10^{-9}$~\cite{Aushev:2010bq} \\
    $\mu \rightarrow e e e$ &  $1.0 \times 10^{-12}$~\cite{Bellgardt:1987du} &  $\sim10^{-16}$~\cite{Blondel:2013ia} \\
    $\tau \rightarrow \mu \mu \mu$ & $2.1\times10^{-8}$~\cite{Hayasaka:2010np} & $\sim 10^{-9}$~\cite{Aushev:2010bq} \\
    $\tau^- \rightarrow e^- \mu^+ \mu^-$ &  $2.7\times10^{-8}$~\cite{Hayasaka:2010np} & $\sim 10^{-9}$~\cite{Aushev:2010bq} \\
    $\tau^- \rightarrow \mu^- e^+ e^-$ &  $1.8\times10^{-8}$~\cite{Hayasaka:2010np} & $\sim 10^{-9}$~\cite{Aushev:2010bq} \\
    $\tau \rightarrow e e e$ & $2.7\times10^{-8}$~\cite{Hayasaka:2010np} &  $\sim 10^{-9}$~\cite{Aushev:2010bq} \\
    $\mu^-, \mathrm{Ti} \rightarrow e^-, \mathrm{Ti}$ &  $4.3\times 10^{-12}$~\cite{Dohmen:1993mp} & $\sim10^{-18}$~\cite{PRIME} \\
    $\mu^-, \mathrm{Au} \rightarrow e^-, \mathrm{Au}$ & $7\times 10^{-13}$~\cite{Bertl:2006up} & \\
    $\mu^-, \mathrm{Al} \rightarrow e^-, \mathrm{Al}$ &  & $10^{-15}-10^{-18}$ \\
    $\mu^-, \mathrm{SiC} \rightarrow e^-, \mathrm{SiC}$ &  & $10^{-14}$~\cite{Natori:2014yba} \\
\hline
\end{tabular}
\caption{Current experimental bounds and future sensitivities for the most important LFV observables.}
\label{tab:sensi}
\end{table}

\subsection{Approximate expressions for the observables}
\label{subsec:approxLFV}

We use the {\tt FlavorKit} \cite{Porod:2014xia} functionality of {\tt
  SARAH}
\cite{Staub:2008uz,Staub:2009bi,Staub:2010jh,Staub:2012pb,Staub:2013tta}
for the analytical computation of the LFV Wilson coefficients and
observables. This allows us to automatically obtain complete
analytical results for the LFV observables as well as robust numerical
routines to be combined with {\tt SPheno}
\cite{Porod:2003um,Porod:2011nf}. For the conventions used in this
paper, the definition of the relevant LFV operators and the generic
expressions for the LFV observables we refer to Appendices
\ref{app:LFVlag} and \ref{app:LFVobs}. Even though we will make use of
the complete analytical results for the numerical exploration of the
phenomenology of the model, we find it convenient to present simple
approximate expressions for the observables of interest.

\begin{figure}[t]
\centering
\includegraphics[width=0.48\textwidth]{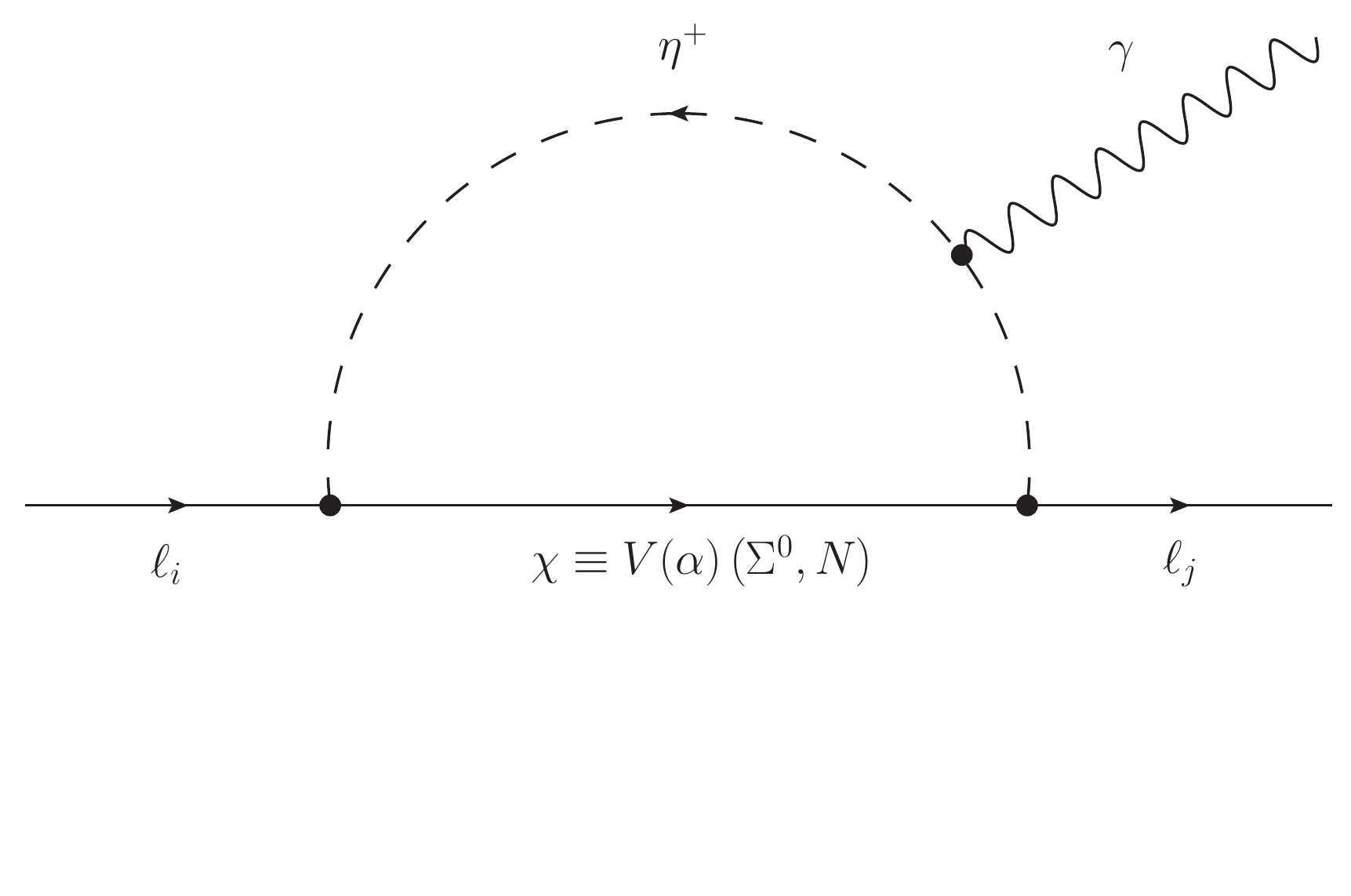}
\includegraphics[width=0.48\textwidth]{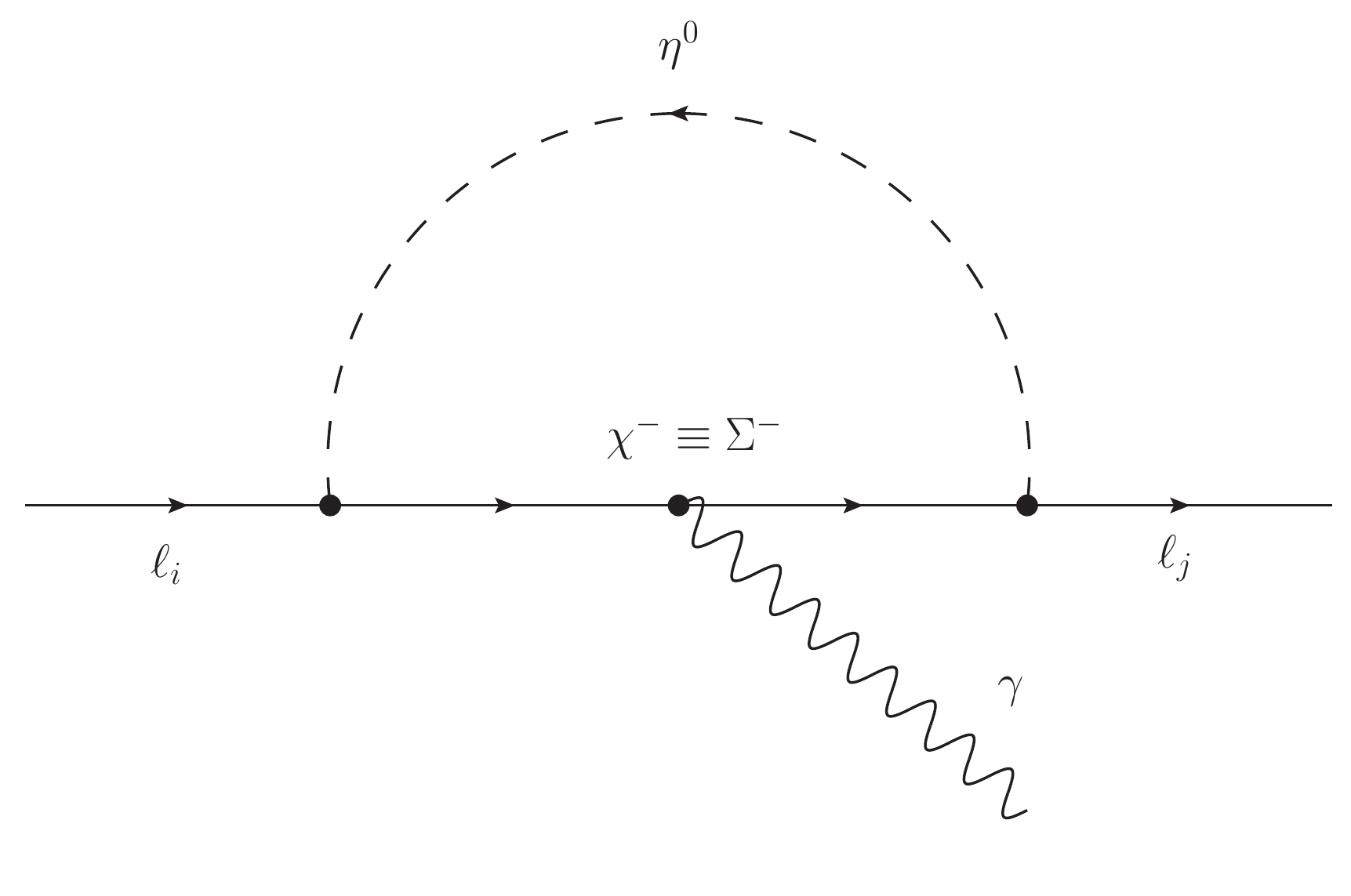}
\caption{Photon penguin diagrams leading to the dominant Wilson
  coefficients $K_1^L$ and $K_2^R$.}
\label{fig:diags}
\end{figure}

Our numerical analysis reveals that the LFV phenomenology is mainly
driven by two Wilson coefficients, both generated by photon penguin
diagrams: the monopole $K_1^L$ and the dipole $K_2^R$. Box diagrams
also lead to sizable contributions, mainly to the Wilson coefficients
$A_{LL}^V$, $B_{LL}^V$ and $C_{LL}^V$, but we have found them to be
always subdominant compared to the photonic monopole and dipole
contributions. Therefore, we can obtain simple approximate expressions
for the LFV observables in terms of only $K_1^L$ and $K_2^R$.

The most relevant photon penguin diagrams in the singlet-triplet
scotogenic model are shown in Fig. \ref{fig:diags}. The diagram with
the neutral fermions $\chi \equiv \left( \chi_1, \chi_2 \right)$
running in the loop is common to the scotogenic model
\cite{Toma:2013zsa}, whereas the diagram with the charged $\Sigma^-$
state is only present in the singlet-triplet variant. This difference
has an impact on the phenomenology, as we will see below. Let us first
consider the dipole coefficient $K_2^R$, which induces the radiative
LFV decay $\ell_\alpha \to \ell_\beta \gamma$. It can be written as
\begin{equation}
K_2^R = \frac{1}{16 \pi^2} \left( D^0 + D^- \right) \, ,
\end{equation}
where the contributions from the two diagrams in Fig. \ref{fig:diags}
are approximately given by
\begin{align}
D^0 =& \, \frac{1}{2 \, m_{\eta^+}^2} \times \nonumber \\
 \bigg[ & \left( \frac{1}{\sqrt{2}}\cos\alpha\sin\alpha\left(\left( Y_\Sigma^i \right)^\ast Y_{N}^j + \left( Y_N^i \right)^\ast Y_{\Sigma}^j \right)+\frac{1}{2}\left( Y_\Sigma^i \right)^\ast Y_{\Sigma}^j \cos^2\alpha+\left( Y_N^i \right)^\ast Y_{N}^j \sin^2\alpha \right) \, F_2(\xi_1) \nonumber \\ 
 + & \left( \frac{-1}{\sqrt{2}}\cos\alpha\sin\alpha\left(\left( Y_\Sigma^i \right)^\ast Y_{N}^j +\left( Y_N^i \right)^\ast Y_{\Sigma}^j \right)+\frac{1}{2} \left( Y_\Sigma^i \right)^\ast Y_{\Sigma}^j \sin^2\alpha+\left( Y_N^i \right)^\ast Y_{N}^j \cos^2\alpha \right) \, F_2(\xi_2) \bigg] \, ,  \label{eq:D0} \\
D^- =& - \frac{1}{2 \, m_{\eta^0}^2} \, \left( Y_\Sigma^i \right)^\ast Y_{\Sigma}^j \, G_2(\rho) \, .  \label{eq:Dm}
\end{align}
Similarly, the monopole coefficient $K_1^L$ can be split as
\begin{equation}
K_1^L = \frac{1}{16 \pi^2} \left( M^0 + M^- \right) \, ,
\end{equation}
and the two contributions from the penguin diagrams in
Fig. \ref{fig:diags} are given by
\begin{align}
M^0 =& \, - \frac{1}{6 \, m_{\eta^+}^2} \times \nonumber \\
 \bigg[ & \left( \frac{1}{\sqrt{2}}\cos\alpha\sin\alpha\left(\left( Y_\Sigma^i \right)^\ast Y_{N}^j + \left( Y_N^i \right)^\ast Y_{\Sigma}^j \right)+\frac{1}{2}\left( Y_\Sigma^i \right)^\ast Y_{\Sigma}^j \cos^2\alpha+\left( Y_N^i \right)^\ast Y_{N}^j \sin^2\alpha \right) \, F_1(\xi_1) \nonumber \\ 
 + & \left( \frac{-1}{\sqrt{2}}\cos\alpha\sin\alpha\left(\left( Y_\Sigma^i \right)^\ast Y_{N}^j +\left( Y_N^i \right)^\ast Y_{\Sigma}^j \right)+\frac{1}{2} \left( Y_\Sigma^i \right)^\ast Y_{\Sigma}^j \sin^2\alpha+\left( Y_N^i \right)^\ast Y_{N}^j \cos^2\alpha \right) \, F_1(\xi_2) \bigg] \, , \label{eq:M0} \\
M^- =& \, \frac{1}{6 \, m_{\eta^0}^2} \, \left( Y_\Sigma^i \right)^\ast Y_{\Sigma}^j \, G_1(\rho) \, . \label{eq:Mm}
\end{align}
Here we have defined
\begin{equation}
\xi_i = \frac{m_{\chi_i}^2}{m_{\eta^+}^2} \, , \quad \rho = \frac{m_{\chi^-}^2}{m_{\eta^0}^2} \, ,
\end{equation}
and used $m_{\eta^R}^2 \simeq m_{\eta^I}^2 \equiv
m_{\eta^0}^2$. Finally, the loop functions appearing in these
expressions are given by
\begin{eqnarray}
F_1(x) &=& \frac{2-9x+18x^2-11x^3+6x^3 \log x}{6(1-x)^4} \, , \\
G_1(x) &=& \frac{-16+45x-36x^2+7x^3+6(3x-2) \log x}{6(1-x)^4} \, , \\
F_2(x) &=& \frac{1-6x+3x^2+2x^3-6x^2 \log x}{6(1-x)^4} \, , \\
G_2(x) &=& \frac{2+3x-6x^2+x^3+6x \log x}{6(1-x)^4} \, .
\end{eqnarray}
We find that in the limit $M_\Sigma \to \infty$ our analytical results
are in good agreement with those obtained in the scotogenic model
\cite{Toma:2013zsa}~\footnote{Notice that the loop functions have been
  renamed with respect to \cite{Toma:2013zsa}.}. Finally, we emphasize
that the numerical results discussed in the next Section are based on
the full 1-loop evaluation of the LFV observables and not on these
approximate expressions, only presented to gain insight.

\section{Phenomenological analysis}
\label{sec:pheno}

Our phenomenological analysis uses a {\tt SARAH}-generated {\tt
  SPheno} \cite{Porod:2003um,Porod:2011nf} module for the numerical
evaluation of the LFV observables. We solve the tadpole equations for
the squared mass terms $m_H^2$ and $m_\Omega^2$ and use an adapted
Casas-Ibarra parameterization for neutrino masses to compute the
Yukawa couplings $Y_N$ and $Y_\Sigma$. For this purpose, the results
of the global fit to neutrino oscillation data \cite{Forero:2014bxa}
will be used. Furthermore, given the little impact on the LFV
phenomenology, we fix the following parameters in the scalar
potential,
\begin{align}
\lambda_{2,3,4} = \lambda_{1,2}^\Omega = \lambda^\eta = 0.1 \, &, \quad \lambda_5 = 10^{-8} \ , \label{eq:par1} \\
\mu_1 = 50 \, \text{GeV} \, &, \quad \mu_2 = 1 \, \text{TeV} \, . \label{eq:par2}
\end{align}
We have explicitly checked that these parameters only affect the LFV
observables indirectly, due to their influence on the scalar
spectrum~\footnote{The parameter $\lambda_5$ does indeed have a strong
  impact on the LFV observables, but only due to the scaling of the
  Yukawa couplings, $Y_N$ and $Y_\Sigma$, induced via the neutrino
  mass relation in Eq. \eqref{eq:mnu}. All our numerical results have
  been obtained with $\lambda_5 = 10^{-8}$, except those for the
  $\tau$ lepton observables, obtained with $\lambda_5 = 10^{-10}$.}.
The large value chosen for the trilinear coupling $\mu_2$ ensures the
conservation of the $\mathbb{Z}_2$ symmetry up to high energy scales
\cite{Merle:2016scw}. We also fix $v_\Omega = 1$ GeV. This choice
leads to a negligible deviation from $\rho = 1$, thus respecting
limits from electroweak precision data.  Finally, the doublet VEV
$v_\phi$ is fixed so that $m_W$ is correctly obtained, see
Eq. \eqref{eq:mW}, and the quartic coupling $\lambda_1$ so that the
lightest CP-even state in the model has a mass compatible with that of
the recently discovered Higgs boson. This leaves us with four free
model parameters,
\begin{equation*}
Y_\Omega \, , \quad m_\eta^2 \, , \quad M_N \, , \quad M_\Sigma \, ,
\end{equation*}
as well as the usual free choices in the implementation of the
Casas-Ibarra parameterization: the $R$ matrix angle $\gamma$, the
Dirac CP-violating phase $\delta$ and Normal/Inverted Hierarchy for
the light neutrino spectrum.

\subsection*{General predictions of the model}

\begin{figure}[t]
\centering
\includegraphics[width=0.48\textwidth]{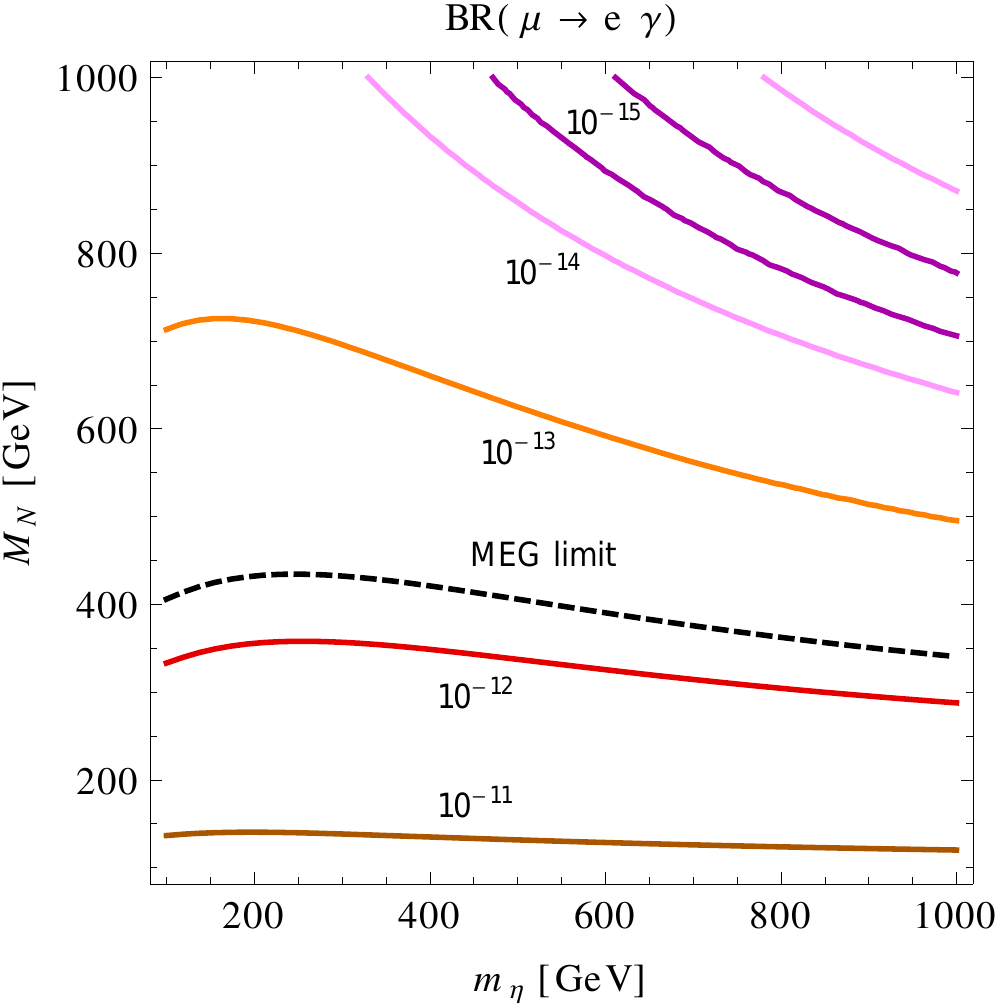}
\includegraphics[width=0.48\textwidth]{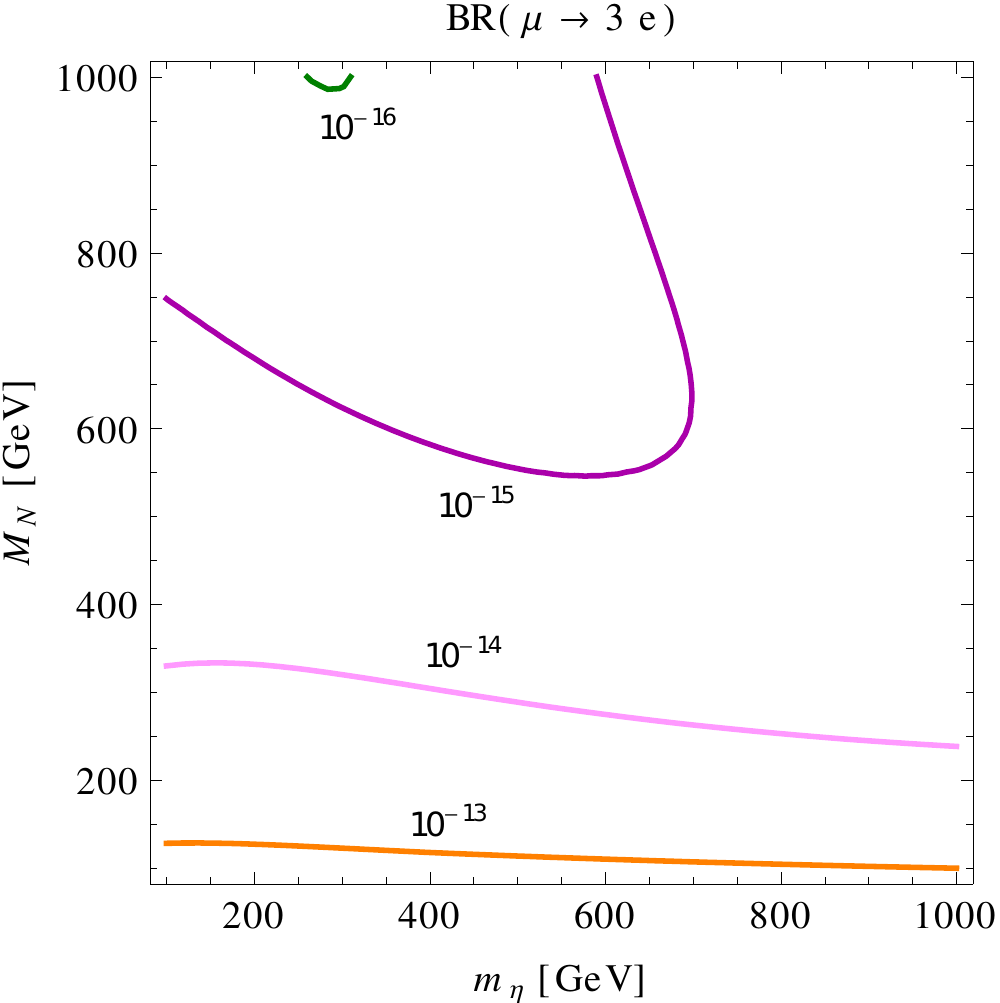} \\
\vspace*{0.5cm}
\includegraphics[width=0.48\textwidth]{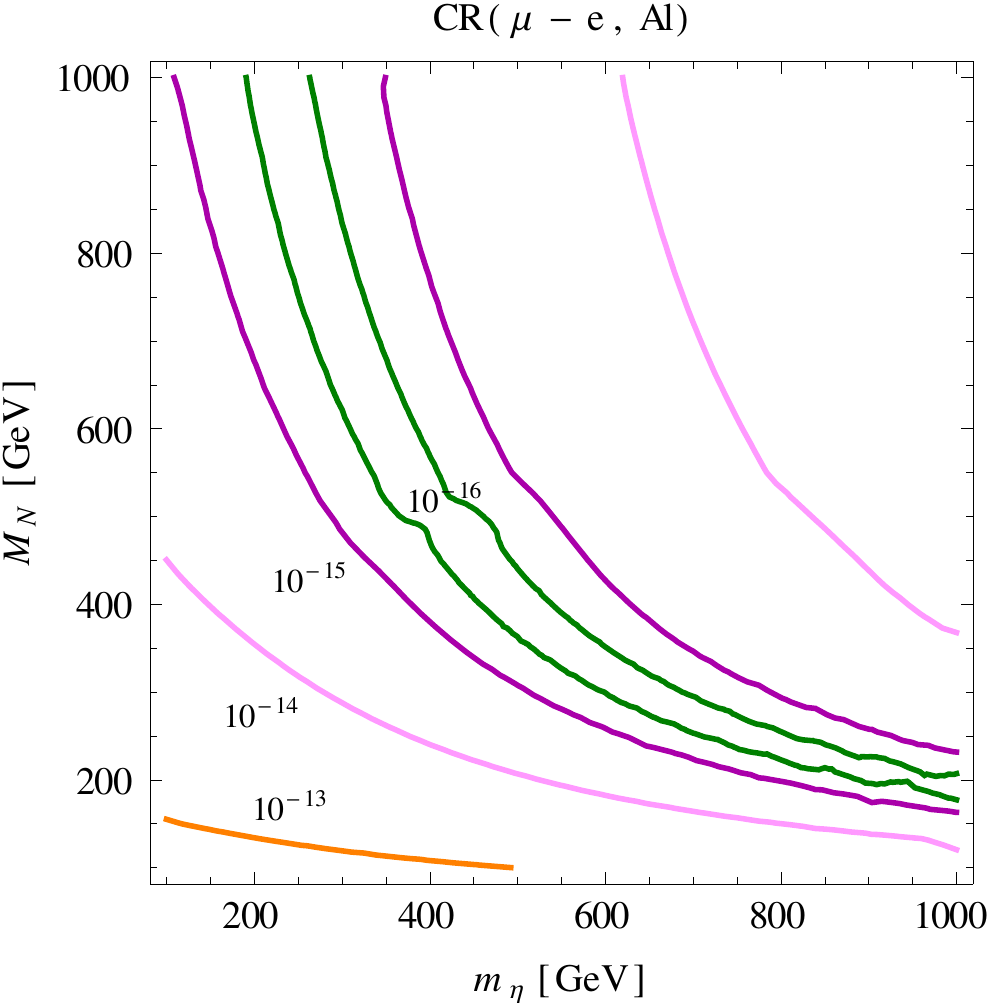}
\caption{Contours of BR($\mu \to e \gamma$), BR($\mu \to 3 \, e$) and
  CR($\mu - e, {\rm Al}$) in the $m_\eta$-$M_N$ plane. Figures
  obtained with fixed $Y_\Omega = 0.1$ and $M_\Sigma = 500$ GeV, see
  text for more details.}
\label{fig:contours}
\end{figure}

We will now explore some aspects of the LFV phenomenology of the
singlet-triplet scotogenic model. First of all,
Fig. \ref{fig:contours} shows contours of BR($\mu \to e \gamma$)
(upper left panel), BR($\mu \to 3 \, e$) (upper right panel) and
CR($\mu - e, {\rm Al}$) (lower panel) in the $m_\eta$-$M_N$ plane,
obtained with the setup introduced above and the choices $Y_\Omega =
0.1$, $M_\Sigma = 500$ GeV, $\gamma = \delta = 0$, normal hierarchy
for the light neutrino spectrum and taking best-fit values for the
neutrino oscillation parameters.  The first conclusion one can draw
from this figure is that the singlet-triplet scotogenic model will be
probed in the next round of LFV experiments: one easily finds
parameter points where the three observables, BR($\mu \to e \gamma$),
BR($\mu \to 3 \, e$) and CR($\mu - e, {\rm Al}$), are within the reach
of the MEG and Mu3e experiments, respectively. In fact, the particular
choice of parameters made in this figure rules out low $M_N$ values
($\lesssim 400$ GeV) as they would imply a too large $\mu \to e
\gamma$ rate, in conflict with the current bound set by the MEG
experiment. In the case of $\mu \to 3 \, e$, the spectacular Mu3e
sensitivity to branching ratios as low as $\sim 10^{-16}$ would allow
one to probe the complete $m_\eta$-$M_N$ plane explored in
Fig. \ref{fig:contours}, with mass values up to the TeV scale and even
higher in some cases. This also happens for $\mu-e$ conversion in
Aluminum.  In this observable, however, a strong cancellation takes
place for a narrow band of the $m_\eta$-$M_N$ plane, where the
resulting negligible conversion rates cannot be probed in the near
future. Qualitatively similar results are found for $\mu-e$ conversion
rates in other nuclei, where analogous cancellations take place as
well.

Figure \ref{fig:contours} also shows that in the long term the
processes $\mu \to 3 \, e$ and $\mu-e$ conversion in nuclei will be
more stringent than $\mu \to e \gamma$. Currently, only the MEG
experiment sets relevant constraints in the explored $m_\eta$-$M_N$,
ruling out a small portion with low $M_N$ values, while the current
bounds for $\mu \to 3 \, e$ and $\mu-e$ conversion in nuclei do not
imply any relevant restrictions. Given the expected experimental
sensitivities in the search for these two observables, this fact will
certainly change in the future. We find that the reach of experiments
such as Mu3e (in case of $\mu \to 3 \, e$) and Mu2e or COMET (in case
of $\mu-e$ conversion in nuclei), clearly supersedes that of MEG, even
after the planned upgrade.

Before moving to the discussion of the BR($\mu \to e \gamma$)/BR($\mu
\to 3 \, e$) ratio, we would like to make some additional comments
about Figure \ref{fig:contours}. We have explicitly checked that our
numerical results reproduce the expected decoupling behavior, namely
that all LFV observables go to zero when $m_\eta$ and $M_{N,\Sigma}$,
the masses of the particles involved in their generation, go to
infinity. However, this is not completely apparent when looking at
Figure \ref{fig:contours}. There are two reasons for this: (i) some
regions of parameter space lead to cancellations among diagrams that
strongly reduce some of the Wilson coefficients (see below for
details), and (ii) the fit to neutrino oscillation data that leads to
an increase in the Yukawa couplings when $m_\eta$ or $M_{N,\Sigma}$
increase.

\subsection*{The $\text{BR}\boldsymbol{(\mu \to 3 \, e)}\boldsymbol{/}\text{BR}\boldsymbol{(\mu \to e \gamma)}$ ratio}

\begin{table}
\centering
\begin{tabular}{c|cc}
\hline
\hline
 & {\bf Point 1} & {\bf Point 2} \\
\hline
$Y_\Omega$ & $0.1$ & $0.1$ \\
$m_\eta^2$ [GeV$^2$] & $2.5 \cdot 10^5$ & $2.5 \cdot 10^{5}$ \\
$M_N$ [GeV] & $500$ & $500$ \\
$M_\Sigma$ [GeV] & $800$ & $300$ \\
\hline
BR($\mu \to e \gamma$) & $4.7 \cdot 10^{-13}$ & $1.3 \cdot 10^{-15}$ \\
BR($\mu \to 3 \, e$) & $3.2 \cdot 10^{-15}$ & $6.1 \cdot 10^{-15}$ \\
CR($\mu- e, {\rm Al}$) & $1.1 \cdot 10^{-15}$ & $5.4 \cdot 10^{-14}$ \\
\hline
\hline
\end{tabular}
\caption{Benchmark points, parameter values and LFV observables. In
  addition to the four input values in this table, we take the
  parameter choices in Eqs. \eqref{eq:par1} and \eqref{eq:par2}, use
  $\gamma = 0$, best-fit values for the neutrino oscillation
  parameters, as obtained in \cite{Forero:2014bxa}, normal hierarchy
  for the light neutrino spectrum and $\delta = 0$. }
\label{tab:benchmark}
\end{table}

We also observe in Fig. \ref{fig:contours} that for most points in the
selected $m_\eta$-$M_N$ plane, one obtains BR($\mu \to e \gamma$)
$\gg$ BR($\mu \to 3 \, e$). However, this is not a general prediction
of the model, as we proceed to discuss now. Let us consider the
benchmark points in Table \ref{tab:benchmark}. The results for the LFV
observables have been obtained making the same choices as for
Fig. \ref{fig:contours}, but using specific values for $m_\eta^2$,
$M_N$ and $M_\Sigma$. First, we observe that the ratio
\begin{equation}
R_{\mu e} = \frac{\text{BR}(\mu \to 3 \, e)}{\text{BR}(\mu \to e \gamma)} \, ,
\end{equation}
can vary by orders of magnitude between different benchmark points
just by changing a single parameter, $M_\Sigma$. In fact, while {\bf
  point 1} predicts LFV rates within the reach of future experiments
searching for $\mu \to e \gamma$, $\mu \to 3 \, e$ and $\mu-e$
conversion in nuclei, {\bf point 2} leads to a BR($\mu \to e \gamma$)
below the foreseen MEG sensitivity and can only be probed by $\mu \to
3 \, e$ and $\mu-e$ conversion in nuclei experiments. Moreover, we
note that only BR($\mu \to e \gamma$) varies substantially between
point 1 and point 2, with a decrease of more than two orders of
magnitude, while the other $\mu-e$ flavor violating observables are
slightly larger in point 2.

\begin{figure}[t]
\centering
\includegraphics[width=0.6\textwidth]{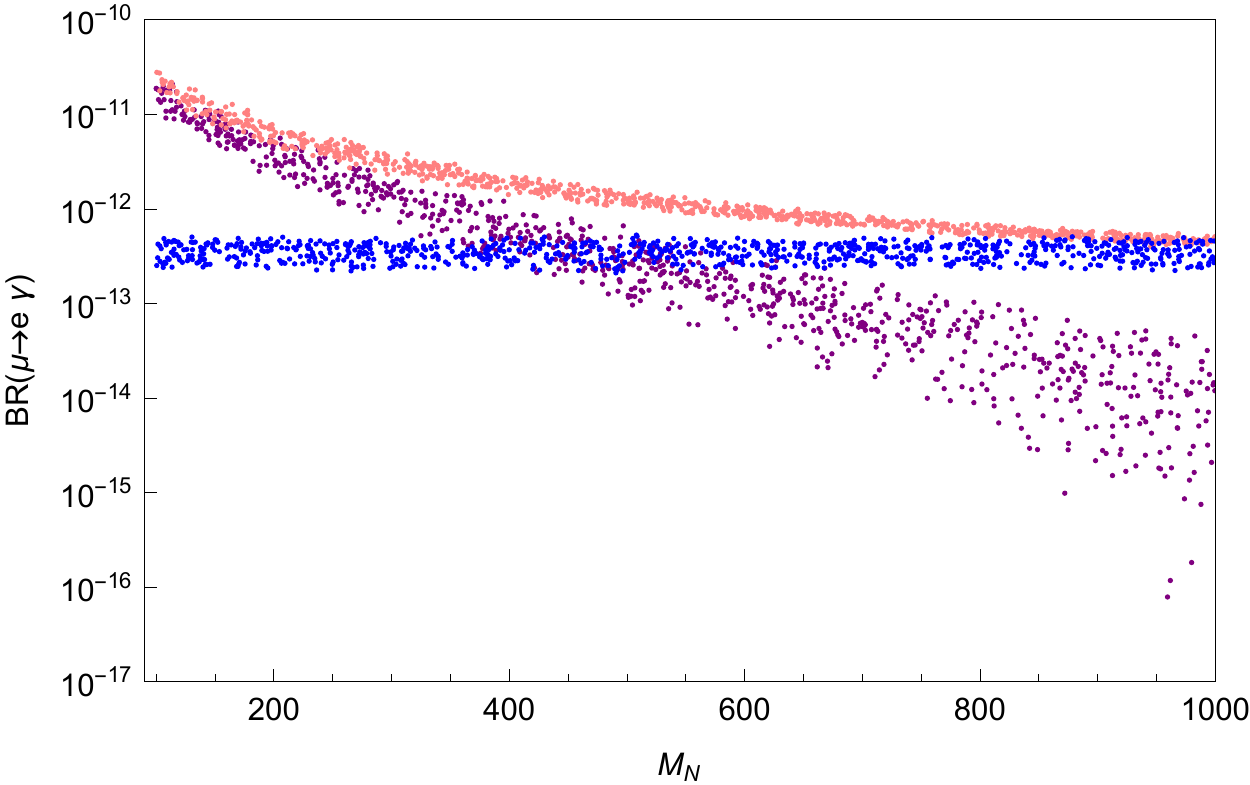}
\caption{BR($\mu \to e \gamma$) as a function of $M_N$ for fixed
  values $Y_\Omega = 0.1$, $m_\eta^2 = 2.5 \cdot 10^5$ GeV$^2$ and
  $M_\Sigma = 500$ GeV. The purple dots display the total branching
  ratio, whereas the pink and blue dots show partial results obtained
  with only the $D^0$ and $D^-$ contributions, respectively.}
\label{fig:cancellation}
\end{figure}

The strong dependence of the $\mu \to e \gamma$ rate on $M_\Sigma$ can
be understood as follows. When $M_\Sigma < M_N$, as in point 2, one
expects the dominant LFV Feynman diagrams to be those with triplet
fermions, $\Sigma^0$ and $\Sigma^-$, running in the loop. Furthermore,
when the mixing between singlet and triplet fermions is small ($\alpha
\simeq 0$) one of the neutral $\chi$ states is mainly composed of
$\Sigma^0$ and is mass degenerate with the charged $\chi^- \equiv
\Sigma^-$. In this case, a cancellation between the $D^0$ and $D^-$
contributions in Eqs. \eqref{eq:D0} and \eqref{eq:Dm} takes
place. Using these equations, it is straightforward to show that for
$\alpha \simeq 0$, the fermion triplet loops lead to $K_2^R \propto
F_2 \left( \xi_1 \right) - 2 \, G_2 \left( \rho \right)$, both loop
functions being positive. Therefore, one naturally expects to find
parameter points where this cancellation in the dipole coefficient is
effective, leading to a reduction in the $\mu \to e \gamma$ rate.

This is explicitly shown in Fig. \ref{fig:cancellation}, where we plot
our numerical results for BR($\mu \to e \gamma$) as a function of
$M_N$ for the fixed values $Y_\Omega = 0.1$, $m_\eta^2 = 2.5 \cdot
10^5$ GeV$^2$ and $M_\Sigma = 500$ GeV. The purple dots display the
total branching ratio, whereas the pink and blue dots show partial
results obtained with only the $D^0$ and $D^-$ contributions,
respectively. This figure has been obtained by allowing the neutrino
oscillation parameters to vary randomly within the preferred $3 \,
\sigma$ ranges found by the global fit of \cite{Forero:2014bxa}, which
explains the spread of the points. We observe that the $D^0$ and $D^-$
contributions approach a common value for large $M_N$ values, whereas
the total branching ratio drops. This is due to the abovementioned
cancellation in the $\Sigma^0$-$\Sigma^-$ loops. For low $M_N$ values
the singlet contributions to $D^0$ dominate and the cancellation in
the triplet contributions is not relevant. However, as $M_N$ increases
and the $N$ contribution to $D^0$ gets smaller, the cancellation in
the triplet contributions becomes visible. We point out that a similar
cancellation in the monopole coefficient takes place, again due to the
relative sign between $M^0$ and $M^-$, see Eqs. \eqref{eq:M0} and
\eqref{eq:Mm}. However, typically this cancellation has little impact
on the LFV observables which receive contributions from the monopole
operator due to the interplay with the other contributions
(e.g. dipole).

\subsection*{LFV $\tau$ decays}

So far we have concentrated on $\mu-e$ violating processes. Now we
turn our attention towards LFV processes involving the $\tau$
lepton. Given the worse experimental limits, these can only be
phenomenologically relevant when they have rates much larger than
those for the $\mu$ lepton. For example, in the benchmark points 1 and
2 presented above one finds branching ratios for the radiative decays
$\tau \to \ell_\alpha \gamma$, with $\ell_\alpha = e, \mu$, in the $\sim
10^{-13}-10^{-12}$ ballpark, clearly below the expected experimental
sensitivity in the near future.

\begin{figure}[t]
\centering
\includegraphics[width=0.48\textwidth]{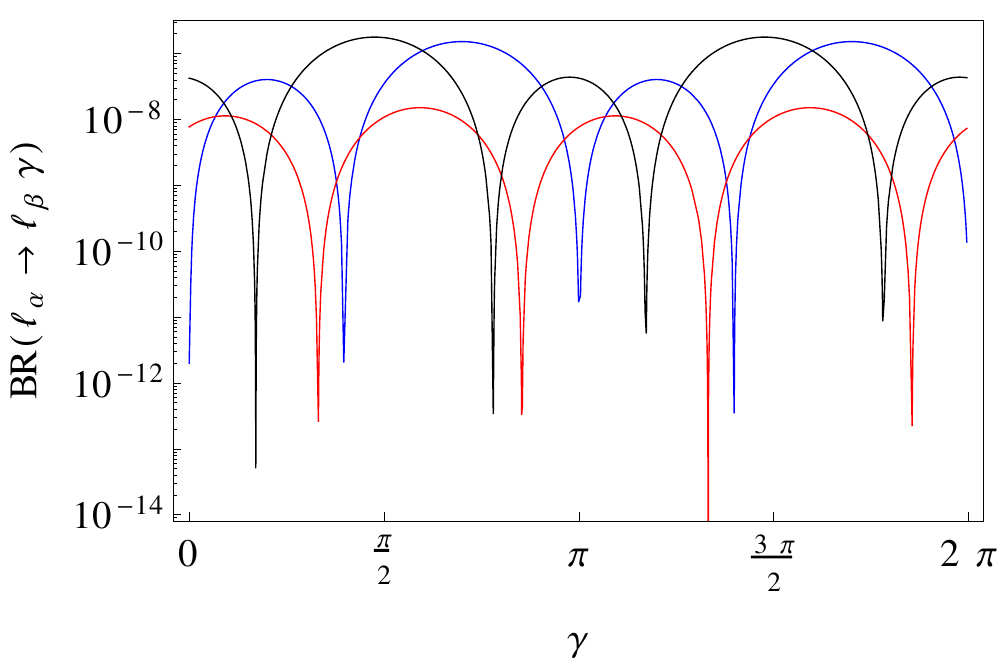}
\includegraphics[width=0.48\textwidth]{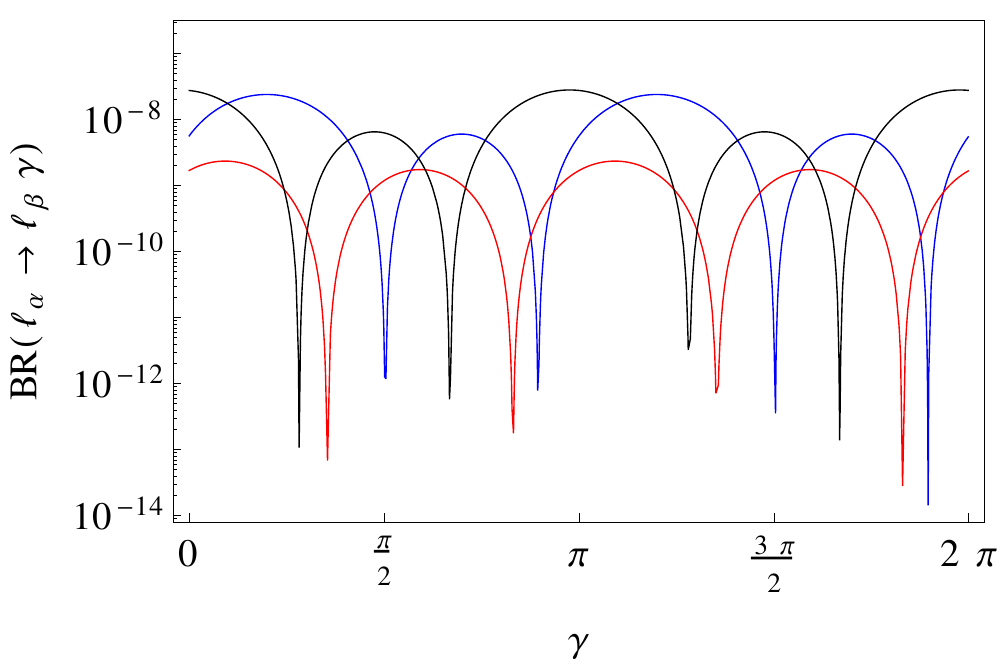}
\caption{BR($\ell_\alpha \to \ell_\beta \gamma$) as a function of the
  $R$ matrix angle $\gamma$ for $M_\Sigma = 300$ GeV (left) and
  $M_\Sigma = 800$ GeV (right). The color code is as follows:
  $(\alpha,\beta) = (2,1)$ in blue, $(\alpha,\beta) = (3,1)$ in red
  and $(\alpha,\beta) = (3,2)$ in black. See text for more details.}
\label{fig:gamma}
\end{figure}

The results shown in Tab. \ref{tab:benchmark} for points 1 and 2 were
obtained with a vanishing $R$ matrix angle $\gamma$. This parameter
has a direct impact on the Yukawa couplings $Y_N$ and $Y_\Sigma$, see
Eqs. \eqref{eq:hfirst} - \eqref{eq:hlast}, and can lead to
cancellations in the amplitudes of specific flavor violating
transitions. This is illustrated in Fig. \ref{fig:gamma}, where we
show our numerical results for BR($\ell_\alpha \to \ell_\beta \gamma$)
as a function of the $R$ matrix angle $\gamma$ (assumed to be real for
simplicity) for $M_\Sigma = 300$ GeV (on the left) and $M_\Sigma =
800$ GeV (on the right). The rest of the parameters are fixed to the
same values as in points 1 and 2, with the exception of a smaller
$\lambda_5$ coupling ($\lambda_5 = 10^{-10}$) in order to increase the
resulting Yukawa couplings and get larger LFV rates. We see in these
figures that even though most points are experimentally excluded due
to a $\mu \to e \gamma$ rate above the MEG bound, for certain $\gamma$
values a strong cancellation takes place, leading to a tiny BR($\mu
\to e \gamma$) and BR($\tau \to e \gamma) \sim 10^{-9}-10^{-8}$
within reach of B factories.

Therefore, we conclude that the singlet-triplet scotogenic model can
also be probed via $\tau$ observables. However, the scenarios that
would be experimentally explored in this way are not generic and
require a certain level of tuning in the Yukawa parameters in order to
suppress the $\mu \to e$ rates.

\section{Summary and conclusions}
\label{sec:conclusions}

We have investigated the lepton flavor violating phenomenology of the
singlet-triplet scotogenic model, a well-motivated scotogenic neutrino
mass model in which neutrinos acquire their masses at the 1-loop
level. The same symmetry that forbids the tree-level generation of
neutrino masses stabilizes a weakly-interacting dark matter candidate,
thus providing a natural solution for another fundamental problem of
current physics.

Our main findings can be summarized as follows:

\begin{itemize}

\item The model will be probed in the next generation of LFV
  experiments. In fact, we have found that parts of the parameter
  space are already ruled out by $\mu \to e \gamma$ searches. This of
  course depends on the value of the $\lambda_5$ parameter, which sets
  the global size of the Yukawa parameters and is expected to be
  naturally small due to its crucial role in the violation of lepton
  number.

\item Currently, the most stringent LFV bound on the model is the one
  set by the MEG experiment on BR($\mu \to e \gamma$). However, this
  will soon change due to the impressive expected sensitivity in the
  incoming experiments. Experiments such as Mu3e (searching for $\mu
  \to 3 \, e$) and Mu2e or COMET (searching for $\mu-e$ conversion in
  nuclei) will soon probe larger portions of the parameter space of
  the model.

\item The operators with the largest contributions to the LFV
  amplitudes are the monopole and dipole ones. These are induced by
  photon penguin diagrams with scotogenic states running in the
  loop. Box diagrams have a subdominant role.

\item One naturally finds points of the parameter space with BR($\mu
  \to 3 \, e$), CR($\mu-e$, Nucleus) $\gg$ BR($\mu \to e
  \gamma$). This is caused by cancellations in the dipole coefficient
  which take place when the dominant contributions are generated by
  $\Sigma^0$-$\Sigma^-$ loops. When this happens, MEG is usually
  unable to constrain the model.

\item The singlet-triplet scotogenic model can also be probed via
  $\tau$ observables, but the scenarios where these have values close
  to the current or near future sensitivities require a certain tuning
  of the Yukawa parameters. Nevertheless, this can be achieved by
  properly choosing the $\gamma$ angle of the Casas-Ibarra matrix $R$.

\end{itemize}

Finally, there are other ways to probe the parameter space of the
singlet-triplet scotogenic model. As already explained, scotogenic
models have a potential interplay between DM physics and LFV in
scenarios with fermionic DM. In this case, the application of LFV
bounds combined with the Planck result for the DM relic density and
contraints from direct DM detection experiments (an attractive feature
of the singlet-triplet scotogenic model), would help obtaining very
stringent constraints on the model and, eventually, ruling out large
fractions of the parameter space. Regarding collider phenomenology,
the $\Sigma$ and $\Omega$ triplets can be pair-produced in Drell-Yan
processes at the LHC. In case of the $\Sigma$ fermions, their
subsequent decays lead to final states including DM particles, hence
to signatures with missing energy, in a way analogous to the standard
R-parity conserving supersimmetric signals
\cite{vonderPahlen:2016cbw}. These interesting possibilities are left
for future work.

\section*{Acknowledgements}

The authors are grateful to J. W. F. Valle and R. Lineros for fruitful
discussions. AV is also grateful to A. Merle, M. Platscher and
N. Rojas for discussions about the singlet-triplet scotogenic model
and collaboration in related projects. Work supported by the Spanish
grants FPA2014-58183-P, Multidark CSD2009-00064, SEV-2014-0398
(MINECO) and PROMETEOII/2014/084 (Generalitat Valenciana). PR was
founded by CONACYT becas en el extranjero CVU 468534. AV acknowledges
financial support from the ``Juan de la Cierva'' program (27-13-463B-
731) funded by the Spanish MINECO.

\appendix

\section{General LFV Lagrangian}
\label{app:LFVlag} 

The general LFV Lagrangian can be split into different pieces
as~\footnote{We closely follow the notation and conventions used in
  {\tt FlavorKit}, see \cite{Porod:2014xia}.}
\begin{equation} \label{eq:L-LFV}
{\cal L}_{\text{LFV}} = {\cal L}_{\ell \ell \gamma} + {\cal L}_{4 \ell} + {\cal L}_{2 \ell 2q} \, .
\end{equation}
The first term is the $\ell - \ell - \gamma$ interaction Lagrangian,
generally given by
\begin{equation} \label{eq:L-llg}
{\cal L}_{\ell \ell \gamma} = e \, \bar \ell_\beta \left[ \gamma^\mu \left(K_1^L P_L + K_1^R P_R \right) + i m_{\ell_\alpha} \sigma^{\mu \nu} q_\nu \left(K_2^L P_L + K_2^R P_R \right) \right] \ell_\alpha A_\mu + \hc
\end{equation}
Here $e$ is the electric charge, $q$ is the photon momentum, $P_{L,R}
= \frac{1}{2} (1 \mp \gamma_5)$ are the usual chirality projectors and
the lepton flavors are denoted by $\ell_{\alpha,\beta}$. We omit
flavor indices in the Wilson coefficients for the sake of clarity. The
first and second terms in Eq.~\eqref{eq:L-llg} are usually called
monopole and dipole operators, respectively. Notice that we have
singled out the photonic contributions, not included in other vector
operators. On the contrary, $Z$- and Higgs boson contributions have
been included whenever possible.

The most general 4-lepton interaction Lagrangian compatible with
Lorentz invariance can be written as
\begin{equation}
{\cal L}_{4 \ell} = \sum_{\substack{I=S,V,T\\X,Y=L,R}} A_{XY}^I \bar \ell_\beta \Gamma_I P_X \ell_\alpha \bar \ell_\delta \Gamma_I P_Y \ell_\gamma + \hc \, , \label{eq:L-4L}
\end{equation}
where we have defined $\Gamma_S = 1$, $\Gamma_V = \gamma_\mu$ and
$\Gamma_T = \sigma_{\mu \nu}$ and $\ell_{\alpha,\beta,\gamma,\delta}$
denote the lepton flavors. Finally, the last piece of
Eq.~\eqref{eq:L-LFV} is the general $2 \ell 2 q$ 4-fermion interaction
Lagrangian, given by
\begin{equation}
{\cal L}_{2 \ell 2q} = {\cal L}_{2 \ell 2d} + {\cal L}_{2 \ell 2u} \label{eq:L-2L2Q}
\end{equation}
where
\begin{eqnarray}
{\cal L}_{2 \ell 2d} = 
&& \sum_{\substack{I=S,V,T\\X,Y=L,R}} B_{XY}^I \bar \ell_\beta \Gamma_I P_X \ell_\alpha \bar d_\gamma \Gamma_I P_Y d_\gamma + \hc \label{eq:L-2L2D} \\
{\cal L}_{2 \ell 2u} = && \left. {\cal L}_{2 \ell 2d} \right|_{d \to u, \, B \to C} \label{eq:L-2L2U} \, ,
\end{eqnarray}
and we have used $d_{\gamma}$ to denote the d-quark flavor.

\section{Generic expressions for the LFV observables}
\label{app:LFVobs}

\subsection{$\ell_\alpha \to \ell_\beta \gamma$}
\label{subsec:LFV1}

The radiative decays $\ell_\alpha \to \ell_\beta \gamma$ only receive
contributions from the dipole operators. The decay width is given
by~\cite{Hisano:1995cp}
\begin{equation}
\Gamma \left( \ell_\alpha \to \ell_\beta \gamma \right) = \frac{\alpha m_{\ell_\alpha}^5}{4} \left( |K_2^L|^2 + |K_2^R|^2 \right) \, ,
\end{equation}
where $\alpha$ is the electromagnetic fine structure constant.

\subsection{$\ell_\alpha \to 3 \, \ell_\beta$}
\label{subsec:LFV2}

In this case, in addition to the standard dipole contributions, the
decay width receives contributions from the monopole operators in
Eq. \eqref{eq:L-llg} and from the 4-lepton operators in
Eq. \eqref{eq:L-4L}. The resulting decay width can be written
as~\cite{Porod:2014xia}
\begin{eqnarray}
\Gamma \left( \ell_\alpha \to 3 \ell_\beta \right) &=& \frac{m_{\ell_\alpha}^5}{512 \pi^3} \left[ e^4 \, \left( \left| K_2^L \right|^2 + \left| K_2^R \right|^2 \right) \left( \frac{16}{3} \log{\frac{m_{\ell_\alpha}}{m_{\ell_\beta}}} - \frac{22}{3} \right) \right. \label{L3Lwidth} \\
&+& \frac{1}{24} \left( \left| A_{LL}^S \right|^2 + \left| A_{RR}^S \right|^2 \right) + \frac{1}{12} \left( \left| A_{LR}^S \right|^2 + \left| A_{RL}^S \right|^2 \right) \nonumber \\
&+& \frac{2}{3} \left( \left| \hat A_{LL}^V \right|^2 + \left| \hat A_{RR}^V \right|^2 \right) + \frac{1}{3} \left( \left| \hat A_{LR}^V \right|^2 + \left| \hat A_{RL}^V \right|^2 \right) + 6 \left( \left| A_{LL}^T \right|^2 + \left| A_{RT}^T \right|^2 \right) \nonumber \\
&+& \frac{e^2}{3} \left( K_2^L A_{RL}^{S \ast} + K_2^R A_{LR}^{S \ast} + c.c. \right) - \frac{2 e^2}{3} \left( K_2^L \hat A_{RL}^{V \ast} + K_2^R \hat A_{LR}^{V \ast} + c.c. \right) \nonumber \\
&-& \frac{4 e^2}{3} \left( K_2^L \hat A_{RR}^{V \ast} + K_2^R \hat A_{LL}^{V \ast} + c.c. \right) \nonumber \\
&-& \left. \frac{1}{2} \left( A_{LL}^S A_{LL}^{T \ast} + A_{RR}^S A_{RR}^{T \ast} + c.c. \right) - \frac{1}{6} \left( A_{LR}^S \hat A_{LR}^{V \ast} + A_{RL}^S \hat A_{RL}^{V \ast} + c.c. \right) \right]  \nonumber \, .
\end{eqnarray}
This expression combines the contributions from monopole
operators with those of 4-lepton operators of vectorial type,
\begin{equation}
\hat A_{XY}^V = A_{XY}^V + e^2 K_1^X \qquad \left( X,Y = L,R \right) \, ,
\end{equation}
and neglects the mass of the leptons in the final state, with the
exception of the dipole contributions $K_2^{L,R}$, where an infrared
divergence would otherwise occur due to the presence of a massless
photon propagator.

\subsection{$\mu-e$ conversion in nuclei}
\label{subsec:LFV3}

In coherent $\mu-e$ conversion in nuclei, only the scalar and vector
operators in Eqs. \eqref{eq:L-llg}, \eqref{eq:L-2L2D} and
\eqref{eq:L-2L2U} contribute.  This includes photonic monopole and
dipole operators, supplemented with the standard photon vertices with
the up- and down quarks, as well as $2 \ell 2 q$ 4-fermion
operators. They induce the effective $\mu e q q$ couplings
\begin{eqnarray}
g_{LV(q)} &=& \frac{\sqrt{2}}{G_F} \left[ e^2 Q_q \left( K_1^L - K_2^R \right)- \frac{1}{2} \left( C_{\ell\ell qq}^{VLL} + C_{\ell\ell qq}^{VLR} \right) \right] \, , \\
g_{RV(q)} &=& \left. g_{LV(q)} \right|_{L \to R} \, , \\ 
g_{LS(q)} &=& - \frac{\sqrt{2}}{G_F} \frac{1}{2} \left( C_{\ell\ell qq}^{SLL} + C_{\ell\ell qq}^{SLR} \right) \, , \\
g_{RS(q)} &=& \left. g_{LS(q)} \right|_{L \to R} \, ,
\end{eqnarray}
where $Q_q$ is the quark electric charge ($Q_d = -1/3$, $Q_u = 2/3$)
and $C_{\ell\ell qq}^{IXK} = B_{XY}^K \, \left( C_{XY}^K \right)$ for
d-quarks (u-quarks), with $X = L, R$ and $K = S, V$. These couplings
at the quark level must be \emph{dressed} to obtain the effective
couplings at the nucleon level. One finds
\begin{align}
g_{XK}^{(0)} &= \frac{1}{2} \sum_{q = u,d,s} \left( g_{XK(q)} G_K^{(q,p)} + g_{XK(q)} G_K^{(q,n)} \right)\,, \\
g_{XK}^{(1)} &= \frac{1}{2} \sum_{q = u,d,s} \left( g_{XK(q)} G_K^{(q,p)} - g_{XK(q)} G_K^{(q,n)} \right)\,,
\end{align}
where the $G_K^{(q,p)}$ and $G_K^{(q,n)}$ numerical coefficients were
computed in \cite{Kosmas:2001mv} and given in
\cite{Porod:2014xia}. For an improved calculation of the scalar
coefficients we refer to \cite{Crivellin:2014cta}. Finally, the
conversion rate, normalized to the standard muon capture rate
$\Gamma_{\rm capt}$, is given by \cite{Kuno:1999jp}
\begin{align}
\text{CR} (\mu- e, {\rm Nucleus}) &= 
\frac{p_e \, E_e \, m_\mu^3 \, G_F^2 \, \alpha^3 
\, Z_{\rm eff}^4 \, F_p^2}{8 \, \pi^2 \, Z \, \Gamma_{\rm capt}}  \nonumber \\
&\times \left\{ \left| (Z + N) \left( g_{LV}^{(0)} + g_{LS}^{(0)} \right) + 
(Z - N) \left( g_{LV}^{(1)} + g_{LS}^{(1)} \right) \right|^2 + 
\right. \nonumber \\
& \ \ \ 
 \ \left. \,\, \left| (Z + N) \left( g_{RV}^{(0)} + g_{RS}^{(0)} \right) + 
(Z - N) \left( g_{RV}^{(1)} + g_{RS}^{(1)} \right) \right|^2 \right\} \, .
\end{align}   
$Z$ and $N$ are the number of protons and neutrons in the nucleus
under consideration and $Z_{\rm eff}$ is its effective atomic
charge~\cite{Chiang:1993xz}.  Furthermore, $G_F$ is the Fermi
constant, $F_p$ is the nuclear matrix element and $p_e$ and $E_e$ (
$\simeq m_\mu$) are the momentum and energy of the electron.

\end{document}